\documentclass[pra,aps,twocolumn,superscriptaddress, longbibliography, notitlepage]{revtex4-1}
\usepackage[colorlinks=true, citecolor=blue, urlcolor=blue ]{hyperref}
\usepackage{graphicx}% Include figure files
\usepackage{dcolumn}% Align table columns on decimal point
\usepackage{bm}
\usepackage{amsmath,amsfonts}
\usepackage[english]{babel}
\usepackage{amsthm}
\usepackage{hyperref}
\usepackage{xcolor}
\theoremstyle{plain}

\begin{document}
%\preprint{APS/123-QED}

%\title{Entanglement witnessed by arbitrarily many independent observers\\recycling a single quantum shared state not violating any Bell inequality %classically correlated
%}

\title{Recycled entanglement detection by arbitrarily many sequential and independent pairs of observers}

% Force line breaks with \\
\author{Mahasweta Pandit}
\affiliation{Institute of Theoretical Physics and Astrophysics, Faculty of Mathematics, Physics and Informatics, University of Gdańsk, 80-308 Gdańsk, Poland}
\author{Chirag Srivastava}
\affiliation{Harish-Chandra Research Institute, HBNI, Chhatnag Road, Jhunsi, Allahabad 211 019, India}
%\email{}
\author{Ujjwal Sen}
\affiliation{Harish-Chandra Research Institute, HBNI, Chhatnag Road, Jhunsi, Allahabad 211 019, India}
%\affiliation{
 %Third institution, the second for Charlie Author
%\affiliation{%



\begin{abstract}
We investigate the witnessing of two-qubit entangled states by sequential and independent pairs of observers, with both observers of each pair acting independently on their part of the shared state from spatially separated laboratories, and subsequently passing their qubits to the next pair in the sequence. It has previously been conjectured that not more than one pair of observers can detect
%such recycling cannot be achieved with respect to  
Clauser-Horne-Shimony-Holt ``Bell-nonlocal'' correlations in a similar set-up. This is intriguing since it is possible to have an arbitrarily long sequence of Bell-nonlocal correlations when only a single observer is allowed to share a bipartite state with multiple observers at the other end. It is therefore interesting to ask whether such restrictions are also present when entangled correlations are considered in the scenario of multiple pairs of observers.
We find that a two-qubit entangled state 
%, shared among a sequence of observers on each side, performing weak measurements independently, 
can be used to witness entanglement arbitrarily many times, by pairs of observers, acting sequentially and independently.
%It is previously conjectured in this scenario that not more than one pair of observer can detect
%such recycling cannot be achieved with respect to  
%Clauser-Horne-Shimony-Holt (CHSH) Bell nonlocal correlations. Interestingly, one can witness entanglement arbitrarily many times. 
We prove the statement to be true when the initial pair  of observers in the sequence share any pure entangled state or when they share a state from a class of mixed entangled states. We demonstrate that the phenomenon can also be observed for a certain class of entangled states in which an arbitrarily long sequence of observer pairs witnessing entanglement is reached in the limit of the initial entanglement content tending to a vanishing amount.
%with the number of pairs of the sequential observers.
%such a feature can also be attained using certain entangled states possessing least amount of entanglement.
\end{abstract}

\maketitle
\section{Introduction}
Quantum entanglement \cite{horodecki09,Guhne09} is among the most crucial resources in quantum information processing and communication. It plays a key role in numerous applications, such as quantum cryptography \cite{Ekert91, bb84}, quantum dense coding \cite{Bennett92}, quantum teleportation \cite{Bennett93}, entanglement swapping \cite{Zukowski93, Bose98} and device-independent tasks like key distribution \cite{Mayers98,Barrett05,Acin07}, randomness amplification \cite{Colbeck12},  randomness expansion \cite{Colbeck111,Colbeck112,Pironio10}, etc. It is therefore useful to learn about 
%necessary to perceive knowledge about 
different techniques of characterizing, quantifying, and utilizing entanglement. 
%One universal 
A valuable 
tool to analyze and characterize entanglement exploits a class of functionals called \emph{entanglement witnesses} (EWs) \cite{Guhne09,horodecki09, Sarbicki14}. Given an entangled state, there always exists an EW \cite{Hor96, terhal00, Sanpera02, Guhne09, guhne03}, and for example,
%in the bipartite scenario, 
Bell inequalities \cite{Bell64, Brunner14}, such as the Clauser-Horne-Shimony-Holt (CHSH) Bell inequality \cite{Clauser69}, that detect the so-called ``Bell-nonlocal'' (or ``nonlocal'')  states are
also EWs.
%, also, in turn, enough to certify the presence of entanglement for nonlocal states. \\

In \cite{Silva15}, the authors introduced a scenario where, given an initial bipartite entangled state, a single observer, Alice, owns one subsystem while the other subsystem is passed among multiple sequential and independent observers, Bobs. The task is to obtain a sequential violation of the CHSH Bell inequality. 
%For $n$ successful violations, the initial state is referred to as an $n$-recyclable with respect to the nonlocal correlations.  
Initially, an unbounded number of recycled nonlocal correlations was detected only for ``biased'' measurement strategies \cite{Silva15}.  Later the same 
was attained for a certain family of nonlocal states using a measurement strategy \cite{Brown20} that is unbiased. However, the question whether the same holds for all nonlocal states remains open. Nonlocal states are certainly entangled, and thus this result also answers the question about arbitrary  recyclability of entanglement for Bell-nonlocal states. Recently, it was also observed that it is possible to witness and recycle entanglement an arbitrary number of times using a CHSH Bell-local entangled state
%, having least amount of entanglement 
\cite{srivastava21}. 
%Thus, it becomes important to classify entangled states which can permit such recyclability.
For other theoretical works in the direction of recyclability of entangled and nonlocal correlations, see   \cite{Mal16,Bera18,Das19,Saha19,Maity20,sriv21,roy20,cabello20,ren21,Fei21,Hall21,Cheng21,zhu21,Das21}. For experimental works, see \cite{Schiavon17,Hu18,Vallone20,Choi20,Feng20}.  
%In \cite{srivastava21}, it was shown that it is also possible to witness entanglement an unbounded number of times in this scenario.  It should be noted that \cite{Brown20} indirectly proves this for some of the nonlocal states while \cite{srivastava21} also proves the statement for states that do not violate CHSH inequality. Therefore, the set of $n$-recyclable entangled states is strictly larger than the set of $n$-recyclable states concerning the nonlocal correlations. Recently \cite{Das21} also provides a context of resource theoretic advantage of such recycling. Other related works on sequential entanglement witness can be found in \cite{Bera18,Maity20,sriv21,Vallone20}.\\

\begin{figure}[t]
    \centering
    \includegraphics[width=0.9\columnwidth]{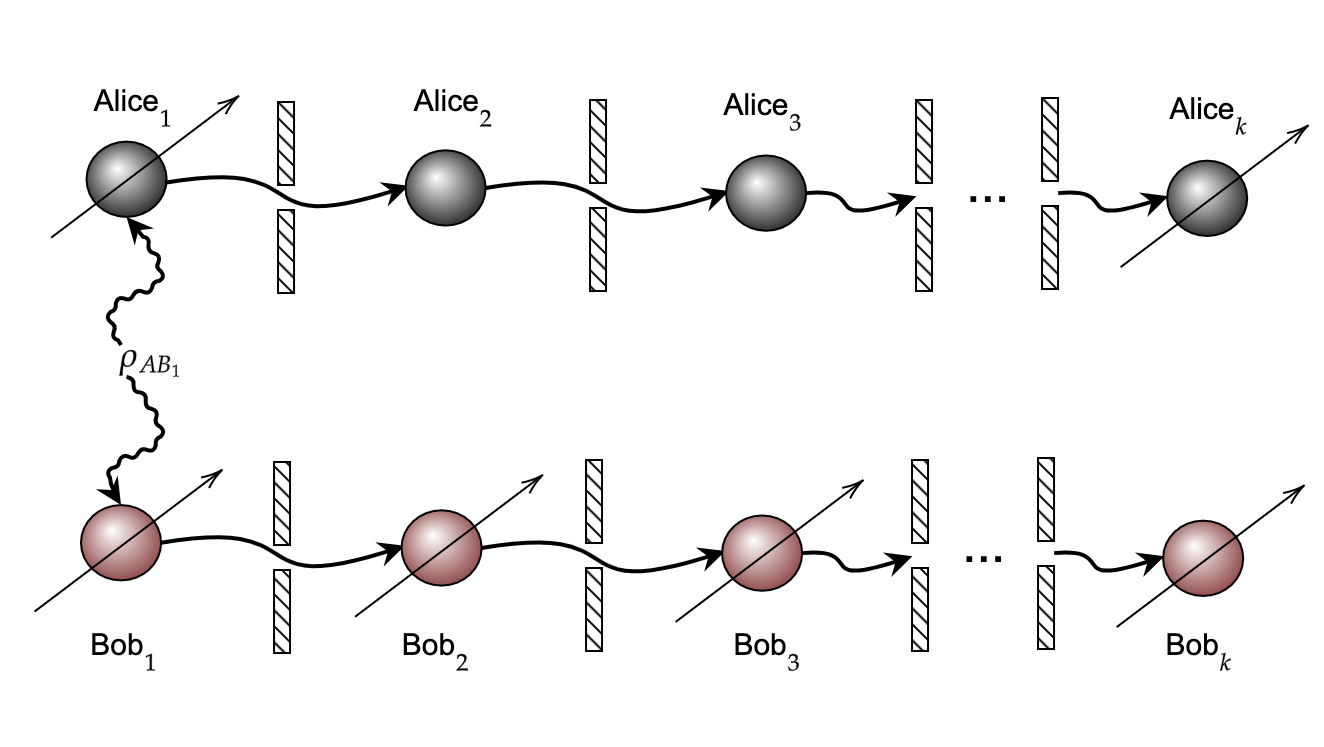}
    \label{plot}
    \caption{Sequentially witnessing entanglement by multiple independent pairs of observers. $\rho_{AB_{1}}$ is the initial state shared between ${\rm Alice}_{1}$ and ${\rm Bob}_{1}$.}
    \label{fig}
\end{figure}

In the present work, we consider the scenario in which both the subsystems of a bipartite quantum state are passed on to multiple pairs of independent observers \cite{Hall21,Cheng21,zhu21,Das21}. See Fig.~\ref{fig}. It has been conjectured in Ref.~\cite{Hall21}, with analytical and numerical evidence, that recycling of nonlocal correlations, via violation of the CHSH Bell inequality, is impossible  in this scenario. That is, not more than one pair of Alice and Bob can share a CHSH Bell nonlocal state. This is interesting, since a single Alice can share CHSH Bell nonlocal correlations with an arbitrary number of sequential and independent Bobs~\cite{Brown20}. But allowing multiple Alices and Bobs seems to restrict such sharability. It is therefore interesting to ask whether entanglement correlations also has such limitations. Contrary to the case of Bell nonlocality, we show that entanglement can be detected and recycled in this scenario, an arbitrary number of times. Therefore, 
%it is interesting that 
where only one pair could detect CHSH Bell nonlocality in  shared two-qubit states, an arbitrary number of pairs can detect entanglement in at least specific instances of the same. We provide a measurement strategy using which, an unbounded number of detections of entanglement can be achieved for any pure entangled two-qubit state and for a class of mixed states. We also show that such a property can be unleashed  by a class of states whose entanglement content vanishes in the limit of the number of observing pairs growing unboundedly. 
%even a weakly entangled state can unleash such property.
%\section{Prerequisites}
%In this section we discuss about the method of entanglement witnesses and unsharp measurements.
\section{Entanglement Witnesses and unsharp measurements}
Entanglement witnesses~\cite{Guhne09,horodecki09,Sarbicki14} use  expectation values of hermitian operators which separate the set of separable states from some of the entangled states. This method of entanglement detection utilizes  the Hahn-Banach theorem~\cite{Simmons63, Lax02}, which says that corresponding to any element falling outside a closed and convex set of a normed linear space, there always exists a functional on that space which ``separates'' the element with the closed and convex set. An entanglement witness operator is thus an operator \(W\), such that 
%Thus, the standard definition of the entanglement witnesses is given as  
$\langle W \rangle_{\rho_s} \geq 0$, for all separable states, $\rho_s$, and there exist at least one entangled state, $\rho_e$, for which  $\langle W \rangle_{\rho_s} < 0$, where $\langle W \rangle_\rho$ represents the expectation value of the hermitian operator, $W$, with respect to a state $\rho$. For example, entanglement of the bipartite state $|\psi^+\rangle=\frac{1}{\sqrt{2}}\left(|01\rangle+|10\rangle\right)$ can be detected by the hermitian operator $|\phi^-\rangle\langle\phi^-|^T$, where $|\phi^-  \rangle=\frac{1}{\sqrt{2}}\left(|00\rangle-|11\rangle\right)$, $T$ represents partial transposition of operators, i.e., transposition with respect to any of the local parties, and $|0\rangle$ and $|1\rangle$ represent the eigenstates of the Pauli operator $\sigma_z$ with eigenvalues  1 and -1, respectively.
The expectation values required for witnessing entangled states can also be computed locally, since hermitian operators acting on a joint Hilbert space can be decomposed in hermitian operators acting on the local Hilbert spaces of the joint Hilbert space. For example, an entanglement witness operator, $W_{\psi^+}$, corresponding to $|\psi^+ \rangle$ is decomposed as~\cite{guhne03}
\begin{equation}
    W_{\psi^+}=\frac{1}{4}\left[ \mathbb{I}_4 +\sigma_z \otimes \sigma_z - \sigma_x \otimes \sigma_x - \sigma_y \otimes \sigma_y \right],
\end{equation}
where $\mathbb{I}_d$ represents identity operator acting on a $d$-dimensional Hilbert space and $\sigma_x$, $\sigma_y$, and $\sigma_z$ are the Pauli matrices.

While the expectation value of $W_{\psi^+}$ will validate the entanglement in $|\psi^+\rangle$, the projective measurements involved to evaluate the expectation value will destroy the entanglement present in the state. But it is possible to detect entanglement as well as preserve some amount of entanglement at the same time, if one performs an unsharp version of the required projective measurements~\cite{Silva15,Brown20}. Assume an observable, $P$, acting on the space of qubits and whose expectation value has to be estimated in some qubit $\rho$.  Let $\{P_0,P_1\}$ denotes the projective measurement such that $P_0+P_1=\mathbb{I}_2$ and $P=P_0-P_1$. This measurement is enough to estimate the required expectation value. One can also define an unsharp version of this measurement for sharpness parameter, $\lambda$, as \begin{eqnarray} \label{kya}
    E^\lambda_0=\frac{1}{2}\left(\mathbb{I}_2+ \lambda P\right), \nonumber\\ 
    E^\lambda_1=\frac{1}{2}\left(\mathbb{I}_2- \lambda P\right),
\end{eqnarray}
where $0\leq\lambda\leq1$.
Note that for $\lambda=1$, the positive operator-valued measure (POVM), $\{E^\lambda_0,E^\lambda_1\}$,  reduces to the projective measurement, $\{P_0,P_1\}$, whereas for $\lambda=0$, measurement operators are identity. Therefore, disturbance in the state is the most when $\lambda=1$ whereas $\lambda=0$ leads to a trivial measurement, in the sense that the state is totally unaffected. 
Notice that the unsharp version of the measurement will lead to an expectaion value of $P$ multiplied by the sharpness parameter, since $ E^\lambda_0- E^\lambda_1= \lambda P$.
The post-measurement state is given by the von Neumann-L\"uder's rule~\cite{Busch} as
\begin{equation}
    \rho \to \sqrt{E^\lambda_0}\rho\sqrt{E^\lambda_0}+\sqrt{E^\lambda_1}\rho\sqrt{E^\lambda_1}.
\end{equation}
\section{Scenario}
We consider a generalized version of the sequential scenario presented in \cite{Silva15}. It involves a bipartite entangled state  that is shared between a pair of observers, namely the $1^{\rm st}$ Alice-Bob pair ($AB_1$). The pair performs their tasks in spatially separated labs. To witness entanglement, the first Alice ($A_1$) and first Bob ($B_1$) each performs one of three local measurements at random on their qubit. Both the post-measurement subsystems are  then passed to $A_2$ and $B_2$ who run the same task on their respective qubits independently and with no prior knowledge of the outcomes attained by their previous observers. This process continues until the state reaches to a pair who fails to detect entanglement. The ultimate aim of this task is to maximize the number of pairs that can witness entanglement independently. We will prove that using the weak measurement strategy presented below, one can achieve an arbitrarily long sequence of Alice-Bob pairs who can successfully detect entanglement for any pure entangled states, a class of mixed entangled states, and weakly entangled states.      
\subsection*{Adopted measurement strategy and entanglement witness}
Let the $k^{\rm th}$ Alice-Bob ($AB_k$) pair share a state $\rho_{AB_k}$ and the witness operator used by the pair is given as follows,
\begin{equation}\label{witness}
    W_k=\frac{1}{4}\left[\mathbb{I}_4 +\sigma_z \otimes \sigma_z - \lambda_k\sigma_x \otimes \lambda_k\sigma_x - \lambda_k\sigma_y \otimes \lambda_k\sigma_y \right],
\end{equation}
where $\lambda_k$ is the sharpness parameter. Therefore, each observer has three measurement settings, i.e., $\{\sigma_x,\sigma_y,\sigma_z\}$. %, and the $\sigma_z$ setting is assumed to be sharp whereas the other settings are unsharp, in general.
For simplicity, we will denote Pauli matrices, $\{\sigma_x,\sigma_y,\sigma_z\}$ as $\{\sigma_1,\sigma_2,\sigma_3\}$ and sharpness parameter, corresponding to any $k^{\rm th}$ local observer and the setting $\sigma_i$ as $\lambda^{(i)}_k$, where $i=1,2,3$. One can see from Eq. \eqref{witness} that 
$$\lambda^{(3)}_k=1~~{\rm and}~~\lambda^{(1)}_k=\lambda^{(2)}_k=\lambda_k.$$
Note that $\langle W_k \rangle \geq 0$ for all separable states as $0\leq \lambda^{(i)}_k\leq 1$~\cite{srivastava21}.
Let $\{\mathcal{A}^{\lambda^{(i)}_k}_0,\mathcal{A}^{\lambda^{(i)}_k}_1\}$ and $\{\mathcal{B}^{\lambda^{(i)}_k}_0,\mathcal{B}^{\lambda^{(i)}_k}_1\}$ denote the measurements performed by the observers $A_k$ and $B_k$ respectively and their forms are taken as given in Eq. \eqref{kya}.
Since it is assumed that all the measurement settings applied by each Alice-Bob pair is equally probable, i.e., unbiased, the state at the hand of the $k^{\rm th}$ pair in sequence is given by
\begin{widetext}
\begin{eqnarray}\label{next}
    \rho_{AB_{k}}&=&\frac{1}{9}\sum_{i,j=1}^3\sum_{a,b=0}^1 \sqrt{\mathcal{A}^{\lambda^{(i)}_{k-1}}_a} \otimes \sqrt{\mathcal{B}^{\lambda^{(j)}_{k-1}}_b} \rho_{AB_{k-1}} \sqrt{\mathcal{A}^{\lambda^{(i)}_{k-1}}_a} \otimes \sqrt{\mathcal{B}^{\lambda^{(j)}_{k-1}}_b}  \nonumber \\
    &=&\frac{1}{36} \sum_{i,j=1}^3\Big[\left(1+\Lambda_{k-1}^{i}\right) \left(1+\Lambda_{k-1}^{j}\right) \rho_{AB_{k-1}}
    + \left(1+\Lambda_{k-1}^{i}\right)\left(1-\Lambda_{k-1}^{j}\right) \mathbb{I}_2\otimes \sigma_j \rho_{AB_{k-1}} \mathbb{I}_2\otimes \sigma_j \nonumber \\
     &+& \left(1-\Lambda_{k-1}^{i}\right)\left(1+\Lambda_{k-1}^{j}\right) \sigma_i \otimes \mathbb{I}_2  \rho_{AB_{k-1}} \sigma_i \otimes \mathbb{I}_2 
      + \left(1-\Lambda_{k-1}^{i}\right)\left(1-\Lambda_{k-1}^{j}\right) \sigma_i \otimes \sigma_j  \rho_{AB_{k-1}} \sigma_i \otimes \sigma_j\Big], \nonumber\\    
\end{eqnarray}
\end{widetext}
where $\Lambda_{k}^{i}=\sqrt{1 - \lambda^{(i)~2}_{k}}$, therefore $\Lambda_{k}^{i}=\sqrt{1 - \lambda^{2}_{k}}$  for $i=1,2$ and will be denoted as  $\Lambda_{k}$.
Now, the expectation values of each term present in the witness operator $W_k$ with respect to $\rho_{AB_k}$ can be expressed in terms of their expectation values with respect to state $\rho_{AB_1}$, i.e.,
\begin{eqnarray}
    &\text{Tr}&[\sigma_z \otimes \sigma_z \rho_{AB_k}]=\text{Tr}[\sigma_z \otimes \sigma_z \rho_{AB_1}]\Pi_{l=1}^{k-1}\frac{\left(1+2\Lambda_l\right)^2}{9}, \nonumber \\
    &\text{Tr}&[\sigma_x \otimes \sigma_x \rho_{AB_k}]=\text{Tr}[\sigma_x \otimes \sigma_x \rho_{AB_1}]\Pi_{l=1}^{k-1}\frac{\left(1+\Lambda_l\right)^2}{9}. \nonumber \\ 
\end{eqnarray}
Note that $\text{Tr}[\sigma_x \otimes \sigma_x \rho_{AB_k}]=\text{Tr}[\sigma_y \otimes \sigma_y \rho_{AB_k}]$ for each $k\in \mathbb{N}.$
\section{Sequential entanglement detection arbitrarily many times}
In this section we study the considered scenario with the first pair of observers in the sequence, sharing classes of state.
\subsection{Maximally entangled state}
Let the pair $AB_1$ shares a maximally entangled state, i.e., $\rho_{AB_1}=|\psi^+\rangle\langle \psi^+|$.
Therefore, the first pair will observe entanglement if ${\rm Tr}[\rho_{AB_1}W_1]<0$,
\begin{equation}
    \implies \lambda^2_1>0.
\end{equation}
And the pair $AB_k$ will able to witness the entanglement if ${\rm Tr}[\rho_{AB_k}W_k]<0$,
\begin{equation}
   \implies \lambda_k^2>\frac{1-\Pi_{l=1}^{k-1}\frac{\left(1+2\Lambda_{l}\right)^2}{9}}
   {2\Pi_{l=1}^{k-1}\frac{\left(1+\Lambda_{l}\right)^2}{9}}.
\end{equation} 
Therefore, let us define the sequence, $\lambda^2_k$, for $k \in \mathbb{N}$ as follows,
\begin{equation}\label{sa}
\lambda^2_k =
\left\{
\begin{array}{cc}
      (1+\epsilon) \frac{1-\Pi_{l=1}^{k-1}\frac{\left(1+2\Lambda_{l}\right)^2}{9}}
   {2\Pi_{l=1}^{k-1}\frac{\left(1+\Lambda_{l}\right)^2}{9}} &\mbox {if } \lambda^2_{k-1}\in (0,1) \\\\
      {\rm \infty}, &{\rm otherwise},
\end{array}
\right.
\end{equation}
with $0<\lambda^2_1<1$ and where $\epsilon > 0$. Note that $\lambda^2_k \in (0,1)$ imply that the pair $AB_k$ will be able to witness the entanglement, whereas $\lambda^2_k = \infty$ implies that the pair $AB_k$ will not be able to witness entanglement. Now for $\lambda^2_k \in (0,1)$,  
\begin{equation}
    \frac{\lambda^2_{k+1}}{\lambda^2_{k}}=\frac{9}{(1+\Lambda_k)^2}\frac{1-\Pi_{l=1}^{k}\frac{\left(1+2\Lambda_{l}\right)^2}{9}}{1-\Pi_{l=1}^{k-1}\frac{\left(1+2\Lambda_{l}\right)^2}{9}} > 1,
\end{equation}
since $\Lambda_k=\sqrt{1-\lambda^2_k} \in (0,1)$. Therefore, the sequence $\lambda^2_k$ in Eq. \eqref{sa} is positive and increasing sequence. The next important observation about this sequence is that as $\lambda^2_1 \to 0$ implies $\lambda^2_k \to 0$ for all $k \in \mathbb{N}$. Thus, it is proved that arbitrary sequence of Alice-Bob pair will able to witness starting from a maximally entangled state.

Interestingly, one can check that the state received by each pair during the process is a CHSH Bell-nonlocal state when the initial state is maximally entangled. Therefore, an arbitrary number of pairs can share Bell nonlocality in the process of witnessing entanglement. However, the measurement settings required to witness such Bell nonlocality is different from the measurement settings that are required to witness the entanglement present in the state. Thus, this observation definitely does not overturn the conjecture made in \cite{Hall21}.
\subsection{Pure entangled states and a class of mixed entangled states}
Any pure entangled state, upto a local unitary, can be written as $|\psi_\alpha\rangle=\sqrt{\alpha}|01\rangle+\sqrt{1-\alpha}|10\rangle$ for $\alpha \in (0,\frac{1}{2}]$, and in Hilbert-Schmidt decomposition can be expressed as 
\begin{equation}
   \begin{split}
     \frac{1}{4}\Big[\mathbb{I}_4 -\sigma_z \otimes \sigma_z &+2\sqrt{\alpha(1-\alpha)} \sigma_x \otimes \sigma_x \\&+ 2\sqrt{\alpha(1-\alpha)}\sigma_y \otimes \sigma_y\Big].
\end{split} 
\end{equation}
Now consider a mixed entangled state shared by $AB_1$,
\begin{equation}
    \rho_{AB_1}=p_1|\psi_\alpha\rangle\langle\psi_\alpha|+p_2|01\rangle\langle01|+p_3|10\rangle\langle10|,
\end{equation}
   where $p_1>0$, $p_2,p_3\geq0$ and $p_1+p_2+p_3=1$. In this case, using the same measurement strategy and witness operator, the sequence of sharpness parameters for the pair of observers will come out to be 
\begin{equation}\label{re}
\lambda^2_k =
\left\{
\begin{array}{cc}
      (1+\epsilon) \frac{1- \Pi_{l=1}^{k-1}\frac{\left(1+2\Lambda_{l}\right)^2}{9}}
   {4p_1\sqrt{\alpha(1-\alpha)}\Pi_{l=1}^{k-1}\frac{\left(1+\Lambda_{l}\right)^2}{9}} &\mbox {if } \lambda^2_{k-1}\in (0,1) \\\\
      {\rm \infty}, &{\rm otherwise},
\end{array}
\right.
\end{equation}
with $0<\lambda^2_1<1$ and where $\epsilon > 0$. This sequence of sharpness parameters is also positive, increasing and goes to zero as $\lambda^2_1 \to 0$. Therefore, an unbounded number of entanglement detection is possible by sequential Alice-Bob pairs where the first sequential pair may share all pure entangled states or particular class of mixed entangled states. 
Note that the unbounded detection of these entangled states can be shown, in similar fashion, even for the scenario where single Alice shares state with multiple Bobs, acting sequentially \cite{Silva15}. But since, it is already known that these states can produce unbounded chain of nonlocal states \cite{Brown20}, which in turn imply entanglement and hence a separate analysis is not needed.

\section{Arbitrary sequence of observers With entanglement of initial state tending to zero}
In the scenario where multiple Bobs could witness and recycle entanglement with a single Alice, an arbitrary number of entanglement detection is possible even with entanglement tending to zero \cite{srivastava21}. While the lesser entanglement imply less amount of resource in various quantum information tasks, nevertheless, it still is a resource and gives quantum advantage in some tasks \cite{Ben93,Hor99,Nielsen02}. Therefore it is also important to investigate whether recycling of states is possible only with highly entangled states or even a weakly entangled states can allow such feat. In this section, we investigate the scenario of multiple pair of observers, detecting and recycling entangled states with the initial pair sharing the state,
\begin{equation}\label{initstate}
\begin{split}
   \rho_{AB_1}=\frac{1}{4}[\mathbb{I}_2 \otimes \mathbb{I}_2 -\cos \theta\sigma_z \otimes \sigma_z + \alpha \sin \theta \sigma_x \otimes \sigma_x\\ + \alpha \sin \theta \sigma_y \otimes \sigma_y], 
\end{split}
\end{equation}
where $\theta \in (0, \frac{\pi}{4}]$ and $\frac{1-\cos\theta}{2\sin\theta}<\alpha\leq 1$. %Entanglement of this state tends to zero as $\theta \to 0$
The $k^{\rm th}$ observer pair, $AB_k$, uses the same witness operator as given in Eq. \eqref{witness}. Therefore, the sequence of the sharpness parameter after applying the condition, ${\rm Tr}[\rho_{AB_k}W_k]<0$, appears as follows:
\begin{equation}\label{ga}
\lambda^2_k =
\left\{
\begin{array}{cc}
      (1+\epsilon) \frac{1-\cos\theta\Pi_{l=1}^{k-1}\frac{\left(1+2\Lambda_{l}\right)^2}{9}}
   {2 \alpha\sin\theta\Pi_{l=1}^{k-1}\frac{\left(1+\Lambda_{l}\right)^2}{9}}, &\mbox {if } \lambda^2_{k-1}\in (0,1) \\\\
      {\rm \infty}, &{\rm otherwise},
\end{array}
\right.
\end{equation}
 with $\lambda^2_1>\frac{1-\cos\theta}{2\alpha\sin\theta}$ and where $\epsilon > 0$. 
  We define another sequence for $\theta \in (0,\frac{\pi}{4}]$,  which upper bounds the sequence in Eq. \eqref{ga}:
\begin{equation}\label{ma}
\gamma^2_k =
\left\{
\begin{array}{cc}
      (1+\epsilon) \frac{1-\left(1-\frac{\theta^2}{2}\right)\Pi_{l=1}^{k-1}\left(1-\frac{2\gamma^2_{k}}{3}\right)^2}
   {\theta\Pi_{l=1}^{k-1}\left(\frac{2-\gamma^2_{k}}{3}\right)^2}, &\mbox {if } \gamma^2_{k-1}\in (0,1) \\\\
      {\rm \infty}, &{\rm otherwise},
\end{array}
\right.
\end{equation}
with $\gamma^2_1=(1+\epsilon)\frac{\theta}{2\alpha}>\lambda^2_1$. It is easy to show that $\gamma^2_k>\lambda^2_k$ for all k such that $\lambda^2_k$ is finite, since $\sqrt{1-x^2}>1-x^2$ for $0<x<1$, $\cos\theta>\left(1-\frac{\theta^2}{2}\right)$, and $\sin\theta>\frac{\theta}{2}$ for $\theta \in (0,\frac{\pi}{4}]$. 
 
Again, similar to the sequences given in Eqs. \eqref{sa} and \eqref{re}, this sequence, $\gamma^2_k$, is also strictly positive and increasing. Also, for $\theta \to 0$ and for $k \in \mathbb{N}$, $\gamma^2_k \to 0$. This implies that $\lambda^2_k \to 0$ as $\theta \to 0$. Therefore, detection and recycling of entanglement is possible when $\theta \to 0$, an arbitrary number of times. It is important to notice that as $\theta \to 0$, entanglement of the state $\rho_{AB_1}$, given in Eq. \eqref{initstate}, tends to zero. Surprisingly therefore, an arbitrary number of pairs of observers is possible even in a situation when the initial state has an entanglement that becomes vanishingly small with the number of pairs growing unboundedly. 
%this is a surprising and interesting result that even starting from a minimal possible amount of entanglement, an arbitrary pairs of observers can witness and recycle entanglement.  

\section{Conclusion}
It is found that an arbitrary number of pairs of sequential and independent observers can witness entanglement, acting locally on two-qubit entangled states. This result is potentially of  importance, given that it has been conjectured that not more than a single pair of observers can detect CHSH Bell-nonlocal correlations in the same scenario~\cite{Hall21,Cheng21,zhu21}. Assuming that the conjecture in the literature is true, our finding demonstrates a  clear distinction between ``entanglement nonlocality'' and ``Bell nonlocality'', with respect to sequential witnessing by pairs of observers.
%fall under different footing under this scenario. 
It is to be noted that in the  case when a single observer share a bipartite quantum state with multiple sequential observers at the other end, entanglement nonlocality and Bell nonlocality behave similarly. There, it is possible to have an arbitrary long sequence for both entangled and Bell-nonlocal correlations \cite{Brown20,srivastava21}.

Here, we identified classes of quantum states which can produce arbitrarily long sequences of entangled correlations. Specifically, all pure entangled states and a certain class of mixed entangled states are helpful in completing the task. It is also shown that one can succeed in such tasks even with the entanglement of initial shared state tending to zero. But whether other mixed entangled states can also produce such arbitrarily long chains remains an open question.

\acknowledgements
The research of CS was supported in part by the INFOSYS scholarship. MP acknowledges the NCN (Poland) grant (grant number 2017/26/E/ST2/01008). The authors from  Harish-Chandra Research Institute acknowledge partial support from the Department of Science and Technology, Government of India through the QuEST grant (grant number DST/ICPS/QUST/Theme-3/2019/120). 

\bibliography{main}

\end{document}